\documentclass[aps,prl,twocolumn,showpacs,showkeys,amsmath,amssymb]{revtex4}
\usepackage{amsfonts}
\usepackage{graphicx}
\usepackage{bm}
\usepackage{graphics,booktabs}
\usepackage{longtable,lscape}
\usepackage{txfonts}
\usepackage{overpic}
\usepackage{amssymb}
\usepackage{epsfig}
\usepackage{feynmf}
\usepackage{epstopdf}
\usepackage{slashed}
\usepackage{multirow}
\usepackage[colorlinks,citecolor=blue,anchorcolor=red,menucolor=red,linkcolor=red,filecolor=red,runcolor=red,urlcolor=blue,frenchlinks=red]{hyperref}

\begin{document}

\title{Establishing low-lying doubly charmed baryons}

\author{Hua-Xing Chen$^1$}
\author{Qiang Mao$^{1,2}$}
\author{Wei Chen$^3$}
\email{chenwei29@mail.sysu.edu.cn}
\author{Xiang Liu$^{4,5}$}
\email{xiangliu@lzu.edu.cn}
\author{Shi-Lin Zhu$^{6,7,8}$}
\email{zhusl@pku.edu.cn}
\affiliation{
$^1$School of Physics and Beijing Key Laboratory of Advanced Nuclear Materials and Physics, Beihang University, Beijing 100191, China \\
$^2$Department of Electrical and Electronic Engineering, Suzhou University, Suzhou 234000, China\\
$^3$School of Physics, Sun Yat-Sen University, Guangzhou 510275, China \\
$^4$School of Physical Science and Technology, Lanzhou University, Lanzhou 730000, China \\
$^5$Research Center for Hadron and CSR Physics, Lanzhou University and Institute of Modern Physics of CAS, Lanzhou 730000, China \\
$^6$School of Physics and State Key Laboratory of Nuclear Physics and Technology, Peking University, Beijing 100871, China \\
$^7$Collaborative Innovation Center of Quantum Matter, Beijing 100871, China \\
$^8$Center of High Energy Physics, Peking University, Beijing 100871, China
}

\begin{abstract}
We systematically study the $S$-wave doubly charmed baryons using the method of QCD sum rules. Our results suggest that the $\Xi_{cc}^{++}$ recently observed by LHCb can be well identified as the $S$-wave $\Xi_{cc}$ state of $J^P = 1/2^+$. We study its relevant $\Omega_{cc}$ state, whose mass is predicted to be around 3.7 GeV. We also systematically study the $P$-wave doubly charmed baryons, whose masses are predicted to be around 4.1 GeV. Especially, there can be several excited doubly charmed baryons in this energy region, and we suggest to search for them in order to study the fine structure of the strong interaction.
\end{abstract}

\maketitle

$\\$
{\it Introduction.}---
Fifteen years ago, the SELEX Collaboration reported an observation of the doubly charmed baryon $\Xi_{cc}^+$ in the $\Xi_{cc}^+\to\Lambda_c^+ K^-\pi^+$ process~\cite{Mattson:2002vu}, and determined its mass to be $3518.9\pm0.9$ MeV~\cite{Olive:2016xmw}. This is the only experimental evidence for the doubly charmed baryons, but all the other experiments did not confirm this signal~\cite{experiment2}, until the recent LHCb experiment~\cite{lhcb}, which observed the $\Xi_{cc}^{++}$ in the $\Lambda_c^+ K^- \pi^+ \pi^+$ invariant mass spectrum. However, its mass was measured to be $3621.40 \pm 0.72 \pm 0.27 \pm 0.14$ MeV by LHCb, which value is significantly larger than that determined by SELEX.
Note that the $\Xi_{cc}^{++}$ and $\Xi_{cc}^{+}$ are isospin partners, and their mass difference should be only a few MeV.

Besides the recent LHCb experiment, there have been many other experiments in recent years~\cite{Olive:2016xmw,experiment3,Aaij:2017nav}, which consequently observed many excited heavy baryons, and gradually make the heavy baryons as well as heavy mesons an ideal platform to study the fine structure of the strong interaction. We refer to the review~\cite{Chen:2016spr} for more discussions on this point.
The doubly charmed baryons have been extensively studied using various theoretical methods, such as various quark models~\cite{quarkmodel}, the bag model~\cite{bagmodel}, QCD sum rules~\cite{sumrule}, lattice QCD simulation~\cite{lattice}, and others (for a incomplete list of works see Refs.~\cite{others}). Again, we refer to Ref.~\cite{Chen:2016spr} for a brief review on these studies. More relevant discussions can be found in Refs.~\cite{review2}.

In this letter we systematically study the $S$-wave and $P$-wave doubly charmed baryons using the method of QCD sum rules~\cite{sumrulereview}. We construct all the local $S$-wave doubly charmed baryon fields by investigating two configurations: one contains a $[cc]$ diquark~\cite{Jaffe:2004ph} together with a light quark, and the other contains a $[cq]$ diquark together with another charm quark. These two configurations can be related by using the Fierz transformation (as long as local fields are used). After carefully examining these relations, we find a doubly charmed baryon field of $J^P = 1/2^+$, where all the three quark fields inside are at the ground-state. We use this field mixed with a few other component to perform QCD sum rule analyses, and find that the $\Xi_{cc}^{++}$ recently observed by LHCb~\cite{lhcb} can be well identified as the $S$-wave $\Xi_{cc}$ state of $J^P = 1/2^+$. 
We also study its relevant $S$-wave $\Omega_{cc}$ state, whose mass is predicted to be around 3.7 GeV.

Following the same approach, we systematically study the $P$-wave doubly charmed baryons, whose masses are predicted to be around 4.1 GeV. Moreover, our results suggest that there can be several excited doubly charmed baryons in this energy region, similar to our previous studies on the excited singly heavy baryons~\cite{oursumrule}, where we also found that there can be several singly excited charmed baryons. Recently, the LHCb experiment observed as many as five excited $\Omega_c$ states~\cite{Aaij:2017nav}. This may also happen for the excited doubly charmed baryons, so we suggest to search for them in the future LHCb and BelleII experiments. We believe this would greatly help to understand the fine structure of the strong interaction, and consequently improve our understanding of the quantum world.

$\\$
{\it Constructions of $S$-wave doubly charmed baryon fields.}---
As the first step, we discuss how we systematically construct the local $S$-wave doubly charmed baryon fields.
They can only have the antisymmetric color configuration $[\epsilon^{abc} q_a c_b c_c]$, where the subscripts $a \cdots c$ are color indices, $q$ represents an up, down or strange quark, and $c$ represents a charm quark. The other structures, including flavor and spin/orbital/total angular momenta, can be generally described by using either
\begin{eqnarray}
\mathcal{B}_1(x) = \epsilon^{abc} \left(c_a^T(x) C \Gamma_1 c_b(x)\right) \Gamma_2 q_c(x)\, ,
\label{def:field1}
\end{eqnarray}
or
\begin{eqnarray}
\mathcal{B}_2(x) = \epsilon^{abc} \left(q_a^T(x) C \Gamma_3 c_b(x)\right) \Gamma_4 c_c(x)\, .
\label{def:field2}
\end{eqnarray}
Here, $C= i\gamma_2 \gamma_0$ is the charge-conjugation operator, the superscript $T$ represents the transpose of the Dirac indices only, and the matrices $\Gamma_{1\cdots4}$ are Dirac matrices describing the Lorentz structure.
The other configuration
\begin{eqnarray}
\mathcal{B}_3(x) = \epsilon^{abc} \left(c_a^T(x) C \Gamma_5 q_b(x)\right) \Gamma_6 c_c(x)\, ,
\label{def:field3}
\end{eqnarray}
can be transformed to be $\mathcal{B}_2(x)$.

The former $\mathcal{B}_1(x)$ contains a $[cc]$ diquark together with a light quark, where we can clearly see the orbital structure between the two charm quarks; while the latter $\mathcal{B}_2(x)$ contains a $[cq]$ diquark together with another charm quark, where we can clearly see the orbital structure between the charm and light quarks inside the $[cq]$ diquark. In the present study we use local fields, so these two configurations can be related by using the Fierz transformation.

The first configuration $\mathcal{B}_1(x)$ can be easily constructed, because we can directly apply the Pauli principle to the two identical charm quarks contained in the $[cc]$ diquark. Following the method used in Ref.~\cite{Chen:2008qv}, we systematically construct all the possible fields, and find altogether two independent Dirac fields (without any free Lorentz index) of the spin-parity $J^P = 1/2^+$, two independent Rarita-Schwinger fields (with one free Lorentz index) of the pure spin-parity $J^P = 3/2^+$, and one tensor field (with two free antisymmetric Lorentz indices) of the same $J^P = 3/2^+$:
\begin{eqnarray}
\eta_1(x) &=& \epsilon^{abc} \left(c_a^T(x) C \gamma_\mu c_b(x)\right) \gamma^\mu \gamma_5 q_c(x) \, ,
\label{def:eta1}
\\
\eta_2(x) &=& \epsilon^{abc} \left(c_a^T(x) C \sigma_{\mu\nu} c_b(x)\right) \sigma^{\mu\nu} \gamma_5 q_c(x) \, ,
\label{def:eta2}
\\
\eta_{3\alpha}(x) &=& \Gamma_{\alpha\mu} \epsilon^{abc} \left(c_a^T(x) C \gamma^{\mu} c_b(x)\right) q_c(x) \, ,
\label{def:eta3}
\\
\eta_{4\alpha}(x) &=& \Gamma_{\alpha\mu} \times \Big( \epsilon^{abc} \left(c_a^T(x) C \sigma^{\mu\nu} c_b(x)\right) \gamma_\nu q_c(x)
 \label{def:eta4}
\\ \nonumber && ~~~ + \epsilon^{abc} \left(c_a^T(x) C \sigma^{\mu\nu} \gamma_5 c_b(x)\right) \gamma_\nu \gamma_5 q_c(x) \Big) \, ,
\\
\eta_{5\alpha_1\alpha_2}(x) &=& \Gamma_{\alpha_1\alpha_2\mu\nu} \times \Big( \epsilon^{abc} \left(c_a^T(x) C \sigma^{\mu\nu} c_b(x)\right) \gamma_5 q_c(x)
\label{def:eta5}
\\ \nonumber && ~~~ + \epsilon^{abc} \left(c_a^T(x) C \sigma^{\mu\nu} \gamma_5 c_b(x)\right) q_c(x) \Big) \, ,
\end{eqnarray}
where $\Gamma_{\alpha\mu}$ and $\Gamma_{\alpha_1\alpha_2\mu\nu}$ are the two projection operators:
\begin{align}
& ~~~ \Gamma_{\mu\nu} = g_{\mu\nu} - {1\over4}\gamma_\mu\gamma_\nu \, ,
\\
& ~~~ \Gamma_{\mu\nu\alpha\beta} = g_{\mu\alpha}g_{\nu\beta} - {1\over2}g_{\nu\beta}\gamma_{\mu}\gamma_\alpha + {1\over2}g_{\mu\beta}\gamma_\nu\gamma_\alpha + {1\over6}\sigma_{\mu\nu}\sigma_{\alpha\beta} \, .
\end{align}
Among them, $\eta_{1,3\alpha}(x)$ contain the $S$-wave $[cc]$ diquark
\begin{eqnarray}
&& \epsilon^{abc} c_a^T(x) C \gamma_\mu c_b(x) ~~~ [^{2S+1}L_J~=~^3S_1] \, ,
\label{def:diquark1}
\end{eqnarray}
while the other three contain excited $[cc]$ diquarks. We can further identify:
\begin{eqnarray}
\nonumber \eta_1(x) &:& s_{[cc]} = 1 \, , \, l_{[cc]} = 0 \, , \, j_{[cc]} = 1 \, , \, J = 1/2 \, ,
\\ \nonumber
\eta_{3\alpha}(x) &:& s_{[cc]} = 1 \, , \, l_{[cc]} = 0 \, , \, j_{[cc]} = 1 \, , \, J = 3/2 \, .
\end{eqnarray}
Here, $s_{[cc]}$, $l_{[cc]}$ and $j_{[cc]}$ are the spin, orbital and total angular momenta of the $[cc]$ diquark, and $J$ is the total angular momentum of the doubly charmed baryon.

The second configuration $\mathcal{B}_2(x)$ can not be so easily constructed.
Still following the method used in Ref.~\cite{Chen:2008qv}, we systematically construct all the possible fields, and find that there are five non-vanishing Dirac fields of $J^P = 1/2^+$:
\begin{eqnarray}
\eta_1^\prime(x) &=& \epsilon^{abc} \left(q_a^T(x) C \gamma_5 c_b(x)\right) c_c(x) \, ,
\label{def:eta1p}
\\
\eta_2^\prime(x) &=& \epsilon^{abc} \left(q_a^T(x) C \gamma_\mu c_b(x)\right) \gamma^\mu \gamma_5 c_c(x) \, ,
\label{def:eta2p}
\\
\eta_6^\prime(x) &=& \epsilon^{abc} \left(q_a^T(x) C c_b(x)\right) \gamma_5 c_c(x) \, ,
\label{def:eta6p}
\\
\eta_7^\prime(x) &=& \epsilon^{abc} \left(q_a^T(x) C \gamma_\mu \gamma_5 c_b(x)\right) \gamma^\mu c_c(x) \, ,
\label{def:eta7p}
\\
\eta_8^\prime(x) &=& \epsilon^{abc} \left(q_a^T(x) C \sigma_{\mu\nu} c_b(x)\right) \sigma^{\mu\nu} \gamma_5 c_c(x) \, .
\label{def:eta8p}
\end{eqnarray}
However, only two of them are independent, and we can use the Fierz transformation to relate them to $\eta_1(x)$ and $\eta_2(x)$:
\begin{eqnarray}
\left( \eta_1^\prime ~~ \eta_2^\prime ~~ \eta_6^\prime ~~ \eta_7^\prime ~~ \eta_8^\prime \right) =
\left( \eta_1 ~~ \eta_2 \right) \times
\left(\begin{array}{ccccc}
-{1\over4} & -{1\over2} & {1\over4}  & {1\over2} & 0
\\
-{1\over8} & 0          & -{1\over8} & 0         & -{1\over2}
\end{array}\right) \, .
\label{eq:fierz}
\end{eqnarray}
Among them, $\eta_{1,2}^\prime(x)$ contain the $S$-wave $[cq]$ diquarks:
\begin{eqnarray}
&& \epsilon^{abc} q_a^T(x) C \gamma_5 c_b(x) ~~~  [^1S_0] \, ,
\label{def:diquark2}
\\ && \epsilon^{abc} q_a^T(x) C \gamma_\mu c_b(x) ~~~ [^3S_1] \, ,
\label{def:diquark3}
\end{eqnarray}
while the other three contain excited $[cq]$ diquarks. We can further identify
\begin{eqnarray}
\nonumber \eta_1^\prime(x) &:& s_{[cq]} = 0 \, , \, l_{[cq]} = 0 \, , \, j_{[cq]} = 0 \, , \, J = 1/2 \, ,
\\ \nonumber
\eta_2^\prime(x) &:& s_{[cq]} = 1 \, , \, l_{[cq]} = 0 \, , \, j_{[cq]} = 1 \, , \, J = 1/2 \, .
\end{eqnarray}
Here, $s_{[cq]}$, $l_{[cq]}$ and $j_{[cq]}$ are the spin, orbital and total angular momenta of the $[cq]$ diquark.
{\it Especially, from Eq.~(\ref{eq:fierz}) we have the relation $\eta_1 = - 2 \eta_{2}^\prime$, making this field interesting because all the three quark fields are at the ground-state.}

Similarly, we find two independent Rarita-Schwinger fields and one tensor field, all of which have $J^P = 3/2^+$:
\begin{eqnarray}
\eta_{3\alpha}^\prime(x) &=& \Gamma_{\alpha\mu} \epsilon^{abc} \left(q_a^T(x) C \gamma^{\mu} c_b(x)\right) c_c(x) \, ,
\label{def:eta3p}
\\
\eta_{4\alpha}^\prime(x) &=& \Gamma_{\alpha\mu} \times \Big( \epsilon^{abc} \left(q_a^T(x) C \sigma^{\mu\nu} c_b(x)\right) \gamma_\nu c_c(x)
\label{def:eta4p}
\\ \nonumber && ~~~ + \epsilon^{abc} \left(q_a^T(x) C \sigma^{\mu\nu} \gamma_5 c_b(x)\right) \gamma_\nu \gamma_5 c_c(x) \Big) \, ,
\\
\eta_{5\alpha_1\alpha_2}^\prime(x) &=& \Gamma_{\alpha_1\alpha_2\mu\nu} \times \Big( \epsilon^{abc} \left(q_a^T(x) C \sigma^{\mu\nu} c_b(x)\right) \gamma_5 c_c(x)
\label{def:eta5p}
\\ \nonumber && ~~~ + \epsilon^{abc} \left(q_a^T(x) C \sigma^{\mu\nu} \gamma_5 c_b(x)\right) c_c(x) \Big) \, .
\end{eqnarray}
The former two fields $\eta_{3\alpha,4\alpha}^\prime(x)$ can be related to $\eta_{3\alpha,4\alpha}(x)$ by using the Fierz transformation,
and $\eta_{5\alpha_1\alpha_2}^\prime(x)$ can also be related to $\eta_{5\alpha_1\alpha_2}(x)$.
Only the first one $\eta_{3\alpha}^\prime(x)$ contains the $S$-wave $[cq]$ diquark, while the other two contain excited diquarks. We can further identify
\begin{eqnarray}
\nonumber \eta_{3\alpha}^\prime(x) &:& s_{[cq]} = 1 \, , \, l_{[cq]} = 0 \, , \, j_{[cq]} = 1 \, , \, J = 3/2 \, .
\end{eqnarray}

$\\$
{\it Interpretations of the $\Xi_{cc}^{++}$.}---
In the present study we shall use $\eta_1 = - 2 \eta_{2}^\prime$, $\eta_{1}^\prime$, $\eta_{3\alpha}$ and $\eta_{3\alpha}^\prime$ to perform QCD sum rule analyses in order to study the $S$-wave doubly charmed baryons of $J^P = 1/2^+$ and $J^P = 3/2^+$. {\it We shall pay special attention to the field $\eta_1(x) = - 2 \eta_{2}^\prime(x)$.} Besides these single fields, we shall also study their mixing
\begin{eqnarray}
\eta^M_1(\theta_1, x) &=& \cos \theta_1 \times \eta_1(x) + \sin \theta_1 \times \eta_{1}^\prime(x) \, ,
\label{def:eta1mix}
\\
\eta^M_{2\alpha}(\theta_2, x) &=& \cos \theta_2 \times \eta_{3\alpha}(x) + \sin \theta_2 \times \eta_{3\alpha}^\prime(x) \, .
\label{def:eta2mix}
\end{eqnarray}
These fields can couple to the doubly charmed baryons $\mathcal{B}$ through
\begin{eqnarray}
\langle 0 | \eta | \mathcal{B} \rangle &=& f_{B} u(p) \, ,
\label{eq:gamma0}
\end{eqnarray}
Then the two-point correlation functions can be written as:
\begin{eqnarray}
\nonumber \Pi\left(q^2\right) = i \int d^4x e^{iq\cdot x} \langle 0 | T\left[\eta(x) \bar \eta(0)\right] | 0 \rangle
= \left( q\!\!\!\slash + M_\mathcal{B} \right) \Pi\left(q^2\right) \, .
\end{eqnarray}
In the present study we use the terms proportional to $M_\mathcal{B}$ to perform numerical analyses. Following Refs.~\cite{Chen:2015ata} we obtain $M_\mathcal{B}$ through:
%
\begin{eqnarray}
M^2_\mathcal{B}(s_0, M_B)
&=& {\int^{s_0}_{s_<} e^{-s/M_B^2} \rho(s) s ds \over \int^{s_0}_{s_<} e^{-s/M_B^2} \rho(s) ds} \, ,
\label{eq:mass}
\end{eqnarray}
%
where $s_<$ is the physical threshold, and $\rho(s)$ is the QCD spectral density which we evaluate up to dimension ten:
\begin{eqnarray}
\rho(s) &=& \rho^{pert}(s) + \rho^{\langle \bar q q \rangle}(s) + \rho^{\langle GG \rangle}(s) + \rho^{\langle \bar q G q \rangle}(s)
\\ \nonumber && ~~~~~~~~~~~ + \rho^{\langle \bar q q \rangle \langle GG \rangle}(s) + \rho^{\langle \bar q G q \rangle \langle GG \rangle}(s) \, ,
\label{eq:ope}
\end{eqnarray}
where $\rho^{pert}(s)$ is the perturbative term; $\rho^{\langle \bar q q \rangle}(s)$, $\rho^{\langle GG \rangle}(s)$, $\rho^{\langle \bar q G q \rangle}(s)$, $\rho^{\langle \bar q q \rangle \langle GG \rangle}(s)$, and $\rho^{\langle \bar q G q \rangle \langle GG \rangle}(s)$ are the terms containing the quark condensate $\langle \bar q q \rangle$, the gluon condensate $\langle g_s^2 GG \rangle$, the quark-gluon mixed condensate $\langle g_s \bar q \sigma G q \rangle$, and their combinations $\langle \bar q q \rangle \langle g_s^2 GG \rangle$ and $\langle g_s \bar q \sigma G q \rangle \langle g_s^2 GG \rangle$, respectively.
We find that the leading perturbative term ($D=0$) and the next-to-leading quark condensate $\langle \bar q q \rangle$ ($D=3$) are important.
The results of these spectral densities are too lengthy, so we list them in the supplementary file ``OPE.nb''.
We use the values listed in Refs.~\cite{Chen:2015ata} for these condensates and the charm quark mass (see also Refs.~\cite{values}).

There are two free parameters in Eq.~(\ref{eq:mass}): the Borel mass $M_B$ and the threshold value $s_0$. In order to obtain reliable QCD sum rule results, we require that the $s_0$ dependence and the $M_B$ dependence of the mass prediction be weak. Beside this, we also need to carefully examine: a) the convergence of the QCD spectral density $\rho(s)$ through:
%
\begin{eqnarray}
{\Pi_8 \over \Pi_{\rm all}} &\equiv& {\int^{\infty}_{s_<} e^{-s/M_B^2} \times \rho^{\langle \bar q q \rangle \langle GG \rangle}(s) \times ds \over \int^{\infty}_{s_<} e^{-s/M_B^2} \times \rho(s) \times ds} \, ,
\label{def:pi8}
\\
{\Pi_{10} \over \Pi_{\rm all}} &\equiv& {\int^{\infty}_{s_<} e^{-s/M_B^2} \times \rho^{\langle \bar q G q \rangle \langle GG \rangle}(s) \times ds \over \int^{\infty}_{s_<} e^{-s/M_B^2} \times \rho(s) \times ds} \, ,
\label{def:pi10}
\end{eqnarray}
%
and b) the pole contribution defined as:
%
\begin{equation}
\mbox{PC} \equiv {\int^{s_0}_{s_<} e^{-s/M_B^2} \times \rho(s) \times ds \over \int^{\infty}_{s_<} e^{-s/M_B^2} \times \rho(s) \times ds} \, .
\label{def:pc}
\end{equation}
%

\begin{figure}[hbtp]
\begin{center}
\begin{tabular}{c}
\scalebox{0.33}{\includegraphics{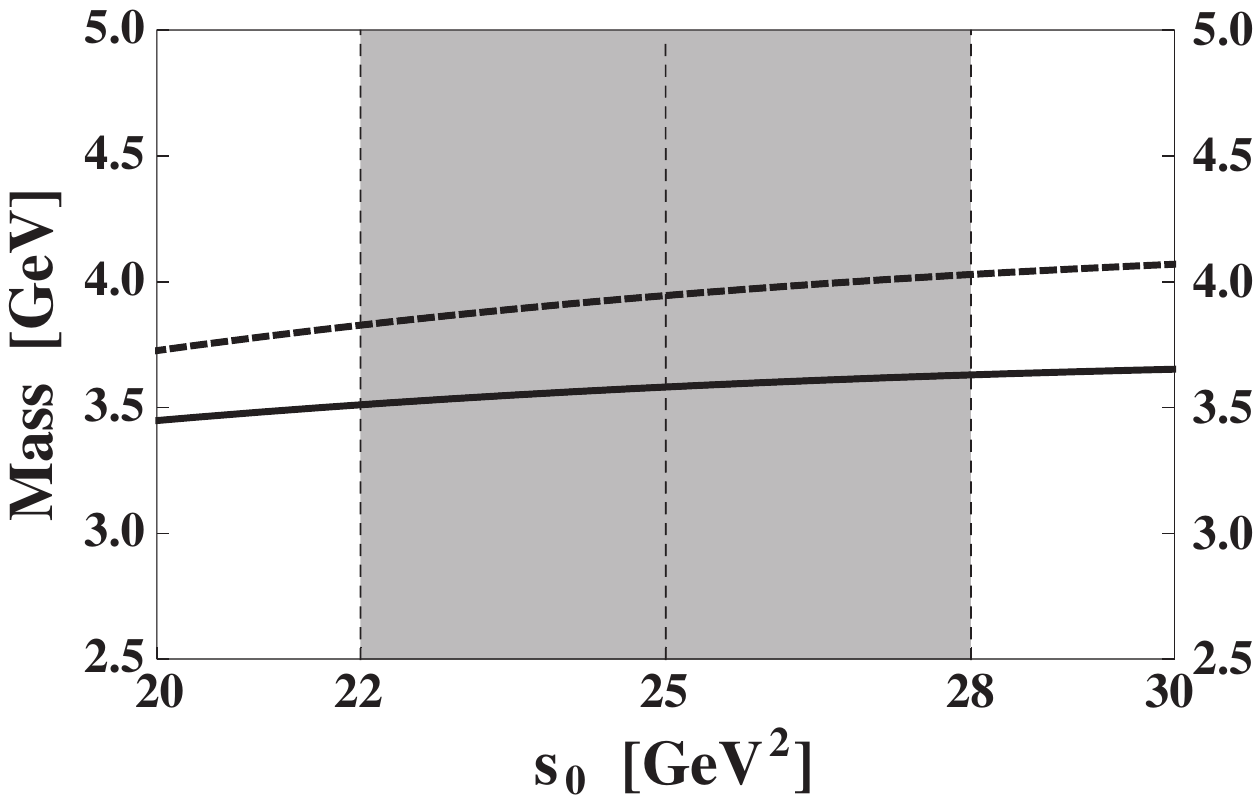}}
\scalebox{0.33}{\includegraphics{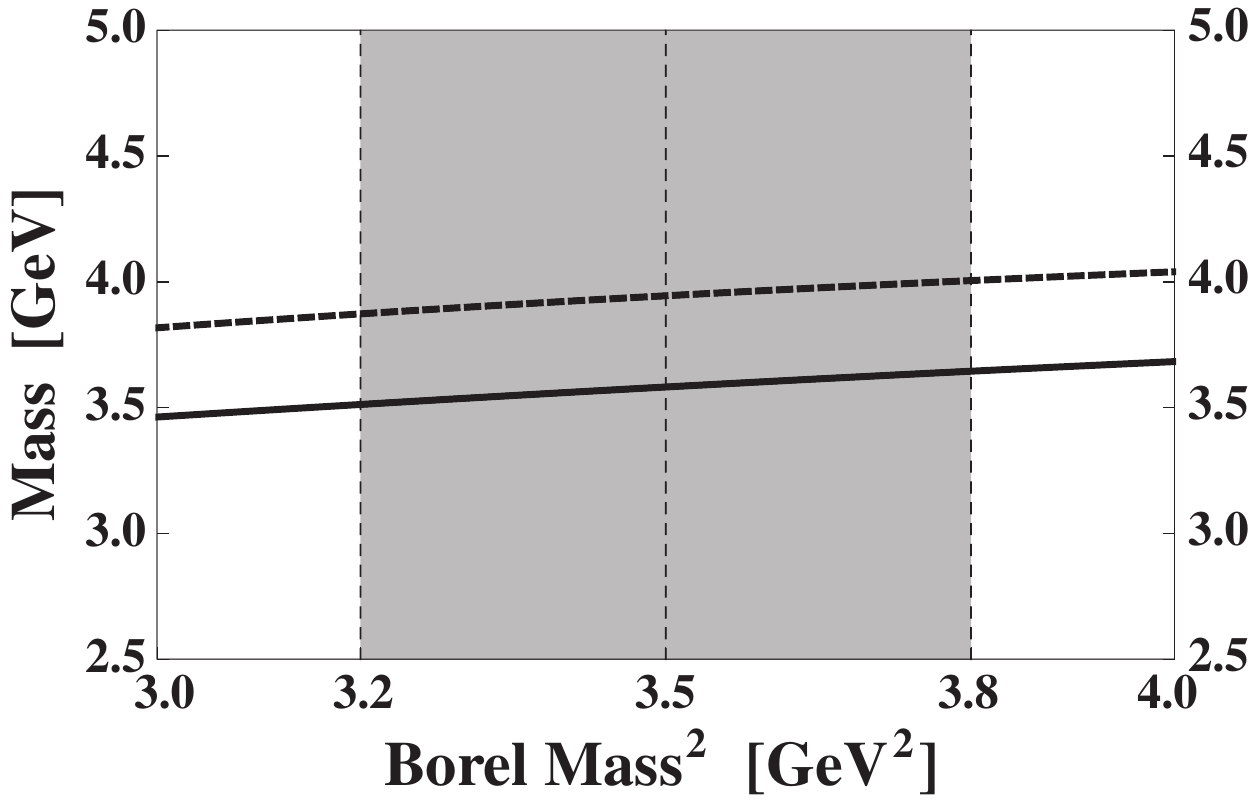}}
\end{tabular}
\caption{Variations of $M_{\eta_{1}^\prime,1/2^+}$ (dashed curves) and $M_{\eta^M_1(\theta_1 = -11^{\rm o}),1/2^+}$ (solid curves) with respect to the threshold value $s_0$ (left) and the Borel mass $M_B$ (right).
}
\label{fig:mass1half}
\end{center}
\end{figure}

\begin{figure}[hbtp]
\begin{center}
\begin{tabular}{c}
\scalebox{0.33}{\includegraphics{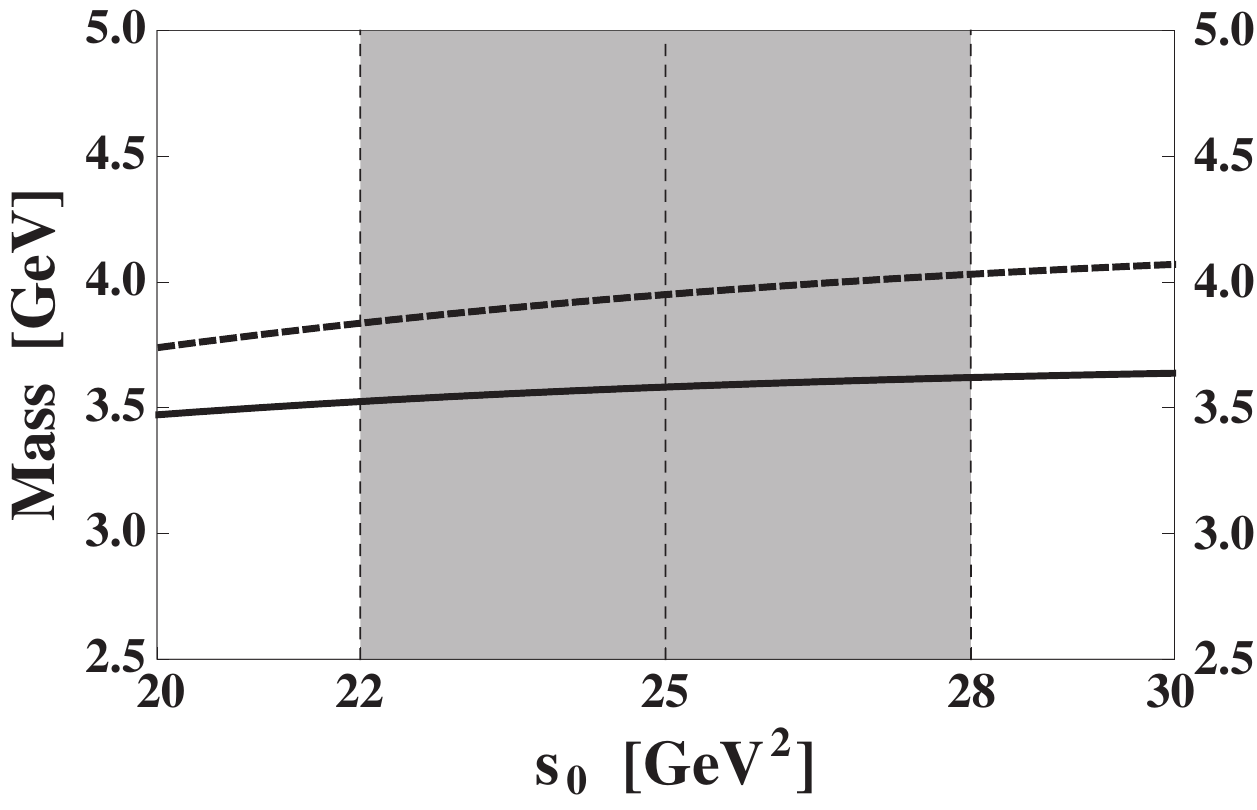}}
\scalebox{0.33}{\includegraphics{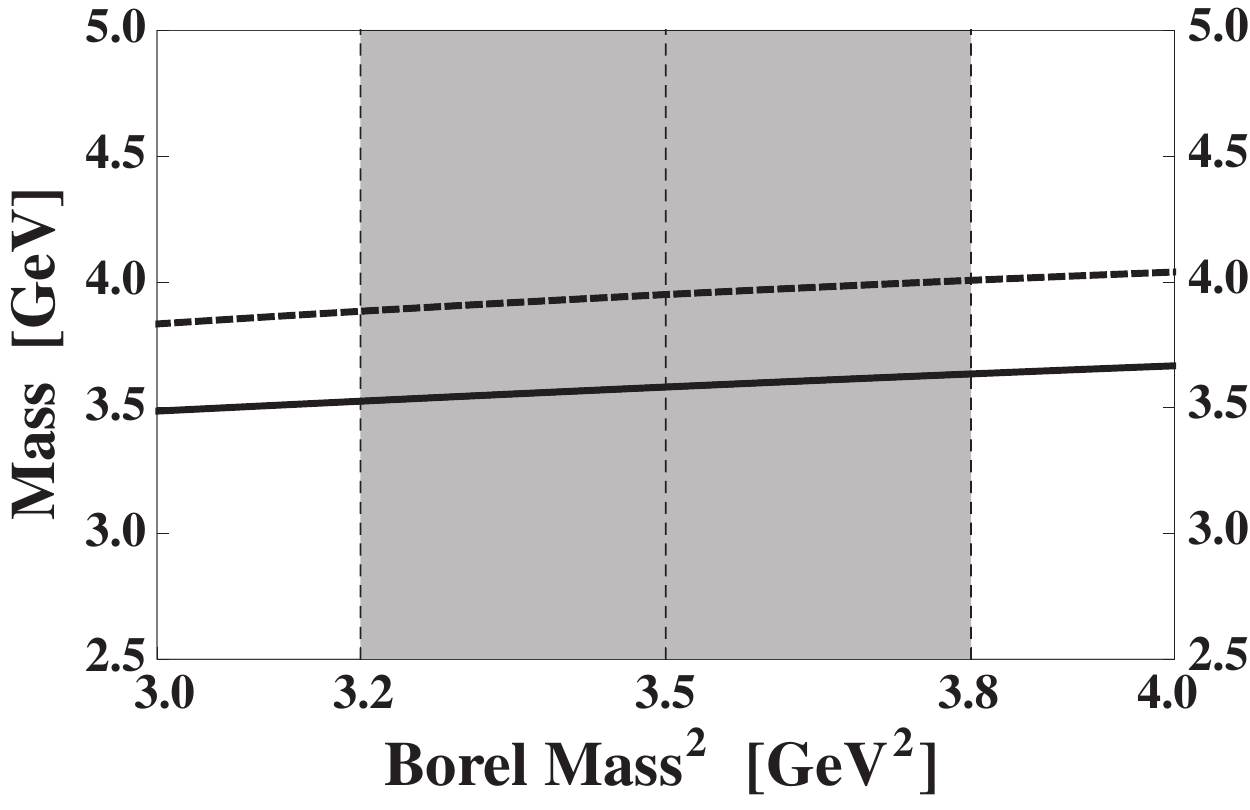}}
\end{tabular}
\caption{Variations of $M_{\eta_{3\alpha}^\prime,3/2^+}$ (dashed curves) and $M_{\eta^M_{2\alpha}(\theta_2 = 6^{\rm o}),3/2^+}$ (solid curves) with respect to the threshold value $s_0$ (left) and the Borel mass $M_B$ (right).
}
\label{fig:mass3half}
\end{center}
\end{figure}

Firstly, we study the $\Xi_{cc}$ by replacing $q \rightarrow u/d$. The masses obtained using $\eta_{1}^\prime$ and $\eta^M_1(\theta_1 = -11^{\rm o})$ are shown in Fig.~\ref{fig:mass1half} as functions of the threshold value $s_0$ and Borel mass $M_B$, and those obtained using $\eta_{3\alpha}^\prime$ and $\eta^M_{2\alpha}(\theta_2 = 6^{\rm o})$ are shown in Fig.~\ref{fig:mass3half}. We choose the working regions of $s_0$ and $M_B$ to be $22$~GeV$^2<s_0<28$~GeV$^2$ and $3.2$~GeV$^2<M_B^2<3.8$~GeV$^2$, and find that the mass curves are quite stable inside these shady regions. We also list them in Table~\ref{tab:results} together with ${\Pi_8 / \Pi_{\rm all}}$, ${\Pi_{10} / \Pi_{\rm all}}$ and PC defined in Eqs.~(\ref{def:pi8}-\ref{def:pc}). We find that the pole contributions are sufficiently large and the OPE spectral densities have good convergence inside these regions, suggesting that our sum rule results are reliable.

\begin{table*}[hbt]
\begin{center}
\caption{Masses and decay constants of the $S$-wave charmed baryons.
}
\begin{tabular}{c | c | c c | c | c c | c c}
\hline\hline
~~\multirow{2}{*}{Fields}~~ & ~~\multirow{2}{*}{Baryons}~~ & ~~\multirow{2}{*}{$s_0$~(GeV)}~ & ~\multirow{2}{*}{$M_B^2$~(GeV$^2$)}~~ & ~~\multirow{2}{*}{Pole}~~ & \multicolumn{2}{c|}{Convergence} & ~~\multirow{2}{*}{Mass~(GeV)}~~ & ~~\multirow{2}{*}{$f$~(GeV$^3$)}~~
\\ \cline{6-7} & & & & & ~~${\Pi_8 / \Pi_{\rm All}}$~ & ~${\Pi_{10} / \Pi_{\rm All}}$~~
\\ \hline
\multirow{2}{*}{$\eta_1=-2\eta_{2}^\prime$}          &
           $\Xi_{cc}$              &  --  &  --  &   --   &   --  &   --   &   --  &   --
\\ &       $\Omega_{cc}$           &  --  &  --  &   --   &   --  &   --   &   --  &   --
\\ \hline
\multirow{2}{*}{$\eta_1^\prime$}          &
           $\Xi_{cc}$              &  $22-28$    &  $3.2-3.8$  &   $>72\%$   &   $<1\%$   &   $<1\%$  &   $3.94^{+0.14}_{-0.15}$  &  $0.12^{+0.03}_{-0.03}$
\\ &       $\Omega_{cc}$           &  $22-28$    &  $3.2-3.8$  &   $>72\%$   &   $<1\%$   &   $<1\%$  &   $3.98^{+0.12}_{-0.16}$  &  $0.13^{+0.02}_{-0.03}$
\\ \hline
\multirow{2}{*}{$\eta_{3\alpha}$}  &
           $\Xi_{cc}$              &  --  &  --  &   --   &   --  &   --   &   --  &   --
\\ &       $\Omega_{cc}$           &  --  &  --  &   --   &   --  &   --   &   --  &   --
\\ \hline
\multirow{2}{*}{$\eta_{3\alpha}^\prime$}  &
           $\Xi_{cc}$              &  $22-28$    &  $3.2-3.8$  &   $>73\%$   &   $<1\%$   &   $<1\%$  &   $3.95^{+0.13}_{-0.15}$  &  $0.11^{+0.01}_{-0.03}$
\\ &       $\Omega_{cc}$           &  $22-28$    &  $3.2-3.8$  &   $>73\%$   &   $<1\%$   &   $<1\%$  &   $3.97^{+0.12}_{-0.17}$  &  $0.11^{+0.02}_{-0.03}$
\\ \hline
\multirow{2}{*}{$\eta^M_1(\theta_1 = -11^{\rm o})$}        &
           $\Xi_{cc}$              &  $22-28$    &  $3.2-3.8$  &   $>87\%$   &   $<1\%$   &   $<1\%$  &   $3.58^{+0.15}_{-0.16}$  &  $0.15^{+0.02}_{-0.03}$
\\ &       $\Omega_{cc}$           &  $22-28$    &  $3.2-3.8$  &   $>84\%$   &   $<1\%$   &   $<1\%$  &   $3.70^{+0.13}_{-0.15}$  &  $0.17^{+0.02}_{-0.03}$
\\ \hline
\multirow{2}{*}{$\eta^M_2(\theta_2 = 6^{\rm o})$}        &
           $\Xi_{cc}$              &  $22-28$    &  $3.2-3.8$  &   $>90\%$   &   $<1\%$   &   $<1\%$  &   $3.58^{+0.14}_{-0.10}$  &  $0.061^{+0.006}_{-0.009}$
\\ &       $\Omega_{cc}$           &  $22-28$    &  $3.2-3.8$  &   $>86\%$   &   $<1\%$   &   $<1\%$  &   $3.69^{+0.11}_{-0.15}$  &  $0.074^{+0.008}_{-0.011}$
\\ \hline \hline
\end{tabular}
\label{tab:results}
\end{center}
\end{table*}

Our results are:
\begin{itemize}

\item The spectral densities extracted from $\eta_1(x) = - 2 \eta_{2}^\prime(x)$ and $\eta_{3\alpha}(x)$ are not complete \Big($\rho^{pert}_{\eta_1}(s) = \rho^{\langle \bar q G q \rangle}_{\eta_1}(s) = 0$ and $\rho^{pert}_{\eta_{3\alpha}}(s) = \rho^{\langle \bar q G q \rangle}_{\eta_{3\alpha}}(s) = 0$\Big), so we can not use them to obtain reliable sum rule results.

\item The masses extracted from $\eta_{1}^\prime(x)$ and $\eta_{3\alpha}^\prime(x)$ are both around 4.0 GeV, significantly larger than the mass of the $\Xi_{cc}$ measured in the SELEX and LHCb experiments.

\item We carefully fine-tune $\theta_1$ to be $-11^{\rm o}$, and use $\eta^M_1(\theta_1 = -11^{\rm o})$ to perform sum rule analyses. The mass is extracted to be
\begin{eqnarray}
M_{\eta^M_1(\theta_1 = -11^{\rm o}),\Xi_{cc}(1/2^+)} &=& 3.58^{+0.15}_{-0.16} {\rm~GeV} \, ,
\end{eqnarray}
where the central value is obtained by choosing $s_0 = 25$ GeV$^2$ and $M_B^2 = 3.5$ GeV$^2$, and the uncertainties come from $\theta_1 \left(= -11\pm5^{\rm o}\right)$, $s_0$, $M_B$, and various quark masses and condensates~\cite{Chen:2015ata}.
Hence, the $\Xi_{cc}^{++}$ recently observed by LHCb~\cite{lhcb} can be well identified as the $S$-wave $\Xi_{cc}$ state of $J^P = 1/2^+$.

\item Similarly, we fine-tune $\theta_2$ to be $6^{\rm o}$, and use $\eta^M_2(\theta_2 = 6^{\rm o})$ to perform sum rule analyses:
\begin{eqnarray}
M_{\eta^M_{2\alpha}(\theta_2 = 6^{\rm o}),\Xi_{cc}(3/2^+)} &=& 3.58^{+0.14}_{-0.10} {\rm~GeV} \, ,
\end{eqnarray}
where the central value is again obtained by choosing $s_0 = 25$ GeV$^2$ and $M_B^2 = 3.5$ GeV$^2$.
Hence, the $\Xi_{cc}^{++}$ may also be identified as the $S$-wave $\Xi_{cc}$ state of $J^P = 3/2^+$.
However, the LHCb experiment preferentially retains longer-lived $\Xi_{cc}^{++}$ candidates~\cite{lhcb}, disfavoring this interpretation because the $\Xi_{cc}$ of $J^P = 3/2^+$ probably has a much shorter lifetime due to its radiative decays.

\item Similarly, we use $\eta^M_1(\theta_1 = -11^{\rm o})$ and $\eta^M_2(\theta_2 = 6^{\rm o})$ to study the $\Omega_{cc}$ by replacing $q \rightarrow s$. The masses are extracted to be
\begin{eqnarray}
M_{\eta^M_1(\theta_1 = -11^{\rm o}),\Omega_{cc}(1/2^+)} &=& 3.70^{+0.13}_{-0.15} {\rm~GeV} \, ,
\\
M_{\eta^M_{2\alpha}(\theta_2 = 6^{\rm o}),\Omega_{cc}(3/2^+)} &=& 3.69^{+0.11}_{-0.15} {\rm~GeV} \, .
\end{eqnarray}
Again, the central values are obtained by choosing $s_0 = 25$ GeV$^2$ and $M_B^2 = 3.5$ GeV$^2$.

\end{itemize}

$\\$
{\it Our study on $P$-wave doubly charmed baryons.}---
The local $P$-wave doubly baryon fields are much more complicated than the $S$-wave ones. We follow Refs.~\cite{oursumrule} and construct them using the $S$-wave diquark fields defined in Eqs.~(\ref{def:diquark1},\ref{def:diquark2}-\ref{def:diquark3}) as well as the following $P$-wave diquark fields:
\begin{eqnarray}
&& \epsilon^{abc} [\mathcal{D}^{\mu} q_a^T(x)] C \gamma_5 c_b(x) ~~~  [^1P_1] \, ,
\\ && \epsilon^{abc} [\mathcal{D}^{\mu} q_a^T(x)] C \gamma^\nu c_b(x) ~~~ [^3P_0]/[^3P_1]/[^3P_2] \, ,
\\ && \epsilon^{abc} [\mathcal{D}^{\mu} c_a^T(x)] C \gamma^\nu c_b(x) ~~~ [^3P_0]/[^3P_1]/[^3P_2] \, ,
\end{eqnarray}
Here, $D^\mu = \partial^\mu + i g A^\mu$ is the gauge-covariant derivative. There are altogether four configurations:
\begin{itemize}

\item Type 1-$[Dq$-$c]c$: we construct three fields of $J^P = 1/2^-$, three of $J^P = 3/2^-$, and one of $J^P = 5/2^-$.

\item Type 2-$[q$-$c]Dc$: we construct three fields of $J^P = 1/2^-$, three of $J^P = 3/2^-$, and one of $J^P = 5/2^-$.

\item Type 3-$[c$-$c]Dq$: we construct two fields of $J^P = 1/2^-$, two of $J^P = 3/2^-$, and one of $J^P = 5/2^-$.

\item Type 4-$[Dc$-$c]q$: we construct one field of $J^P = 1/2^-$ and one of $J^P = 3/2^-$.

\end{itemize}
We note that: a) these exist more non-vanishing and independent fields containing other diquark fields, and b) there are some relations among these configurations due to the Fierz transformation.

We use all these $P$-wave doubly baryon fields to perform QCD sum rule analyses and systematically studied the $P$-wave doubly charmed baryons of $J^P = 1/2^-$, $3/2^-$ and $5/2^-$.
We note that we have only used the single fields but have not investigated their mixing.
We find that both type 1 ($[Dq$-$c]c$) and type 2 ($[q$-$c]Dc$) can lead to reasonable sum rule analyses, and our results suggest that the masses of the $P$-wave doubly charmed baryons (both $\Xi_{cc}$ and $\Omega_{cc}$) are around 4.0-4.2 GeV. Moreover, our results suggest that there can be several excited doubly charmed baryons in this energy region. We shall detailly discuss these results in our further work.

$\\$
{\it Summary and discussions.}---
We have systematically studied the $S$-wave and $P$-wave doubly charmed baryons using the method of QCD sum rules. We construct all the local $S$-wave doubly charmed baryon fields, and find one field of $J^P = 1/2^+$ where all the three quark fields inside are at the ground-state (verified by the Fierz transformation). We use this field mixed with a few other component to perform QCD sum rule analyses, and find that the $\Xi_{cc}^{++}$ recently observed by LHCb~\cite{lhcb} can be well identified as the $S$-wave $\Xi_{cc}$ state of $J^P = 1/2^+$. We have also studied its relevant $S$-wave $\Omega_{cc}$ state, whose mass is predicted to be around 3.7 GeV. We suggest to search for them in the future LHCb and BelleII experiments in the Cabibbo-favored weak decay channels, such as $\Xi^{(\prime)(*)0} D^{(*)+}$, $\Xi_c^{(\prime)(*)+} \bar K^{(*)0}$, and $\Omega_c^{(*)0} \pi^+(\rho^+)$, etc.

Following the same approach, we have systematically studied the $P$-wave doubly charmed baryons, whose masses are predicted to be around 4.1 GeV. Our results suggest that there can be several excited doubly charmed baryons in this energy region, similar to our previous studies on the excited singly heavy baryons~\cite{oursumrule}, where we also found that there can be several singly charmed baryons. Recalling that the LHCb experiment observed as many as five excited $\Omega_c$ states~\cite{Aaij:2017nav}, we also suggest to search for these excited doubly charmed baryons in the future LHCb and BelleII experiments in similar Cabibbo-favored weak decay channels \Big($\Xi_{cc} \to \Sigma^{(*)} D^{(*)}, \Lambda D^{(*)}, \Sigma_c^{(*)} K^{(*)}, \Lambda_c K^{(*)}, \Xi_c^{(\prime)(*)} \pi(\rho)$ and $\Omega_{cc} \to \Xi^{(\prime)(*)} D^{(*)}, \Xi_c^{(\prime)(*)} K^{(*)}, \Omega_c^{(*)} \pi(\rho)$, etc.\Big) in order to study the fine structure of the strong interaction.

\section*{Acknowledgments}

We thank Fu-Sheng Yu for helpful discussions.
This project is supported by the National Natural Science Foundation of China under Grants
No. 11475015, No. 11375024, No. 11222547, No. 11175073, No. 11575008, and No. 11621131001;
the 973 program; the Ministry of Education of China (SRFDP under Grant No. 20120211110002 and the Fundamental Research
Funds for the Central Universities); the National Program for Support of Top-notch Youth Professionals.

\end{document}